\nonstopmode

\newcommand{\ie}{i.e.{}}
\newcommand{\eg}{e.g.{}}
\newcommand{\eV}{\U{eV}}
\newcommand{\Cal}[1]{{\cal #1}}
\newcommand{\etal}{\textit{et al.}}
\newcommand{\U}[1]{\,{\rm{#1}}}
\newcommand{\I}[1]{_{\mathrm{#1}}}
\newcommand{\mat}[1]{\hbox{\boldmath{$#1$}\unboldmath}}
\newcommand{\imag}{{\rm i}}
\newcommand{\euler}{\mathrm e}
\newcommand{\Sum}{\sum\limits}
\newcommand{\Int}{\int\limits}
\newcommand{\unitmatrix}{\mat{\mathbbm{1}}}
\newcommand{\differential}{\>\mathrm d}
\newcommand{\transpose}{{}^{\textrm{\scriptsize T}}}
\newcommand{\bra}[1]{\left<\right.\!#1\!\left.\right|}
\newcommand{\ket}[1]{\left|\right.\!#1\!\left.\right>}
\newcommand{\E}[1]{\times 10^{#1}}
\newcommand{\XUV}{\textsc{xuv}}
\newcommand{\atopa}[2]{\genfrac{}{}{0pt}{}{#1}{#2}}

\documentclass[10pt,letterpaper,twocolumn]{article}
\usepackage{ol2,bbm,amsmath,amssymb,graphicx}
\usepackage[nativepdf,bookmarks,bookmarksopen,bookmarksnumbered,raiselinks,%
breaklinks,%
pdftitle={High-order harmonic generation enhanced by XUV light},%
pdfauthor={Christian Buth, Markus C. Kohler, Joachim Ullrich, %
Christoph H. Keitel},%
pdfsubject={Atomic physics},%
pdfkeywords={XUV, photoabsorption, core-hole decay, krypton, high-order %
harmonic generation, HHG, x-ray free electron laser, FEL, Rabi flopping, %
strong-field physics, optical laser}]{hyperref}

\begin{document}
\twocolumn[ 

\title{High-order harmonic generation enhanced by XUV~light}
\author{Christian Buth,$^{1,2,*}$ Markus C. Kohler,$^1$
Joachim Ullrich,$^{1,3}$ and Christoph H. Keitel$^1$}
\address{
$^1$Max-Planck-Institut f\"ur Kernphysik, Saupfercheckweg~1,
69117~Heidelberg, Germany \\
$^2$Argonne National Laboratory, Argonne, Illinois~60439, USA \\
$^3$Max Planck Advanced Study Group at the Center for Free-Electron
Laser Science, Notkestra\ss{}e~85, 22607~Hamburg, Germany \\
$^*$Corresponding author: christian.buth@web.de
}

\begin{abstract}
The combination of high-order harmonic generation~(HHG) with resonant
\XUV{}~excitation of a core electron into the transient
valence vacancy that is created in the course of the HHG~process is
investigated theoretically.
In this setup, the first electron performs a HHG three-step process
whereas, the second electron Rabi flops
between the core and the valence vacancy.
The modified HHG~spectrum due to recombination with the valence \emph{and}
the core is determined and analyzed for krypton on
the~$3d \to 4p$~resonance in the ion.
We assume an $800 \U{nm}$~laser with an intensity of about~$10^{14}
\U{\frac{W}{cm^2}}$ and \XUV{}~radiation from the Free Electron Laser
in Hamburg~(FLASH) with an intensity in the
range~$10^{13}$--$10^{16} \U{\frac{W}{cm^2}}$.
Our prediction opens perspectives for nonlinear \XUV{}~physics,
attosecond x~rays, and HHG-based spectroscopy involving core orbitals.
\end{abstract}

%
%
%
%
%
%


] 

\renewcommand{\onlinecite}[1]{\cite{#1}}

High-order harmonic generation~(HHG) by atoms in intense optical laser fields
is a fascinating phenomenon and a versatile tool;
it has spawned the field of attoscience, is used for spectroscopy,
and serves as a light source in many optical
laboratories~\cite{Agostini:PA-04}.
Present-day theory of HHG largely gravitates around the
single-active electron~(SAE) approximation and the
restriction to HHG from valence electrons~\cite{Schafer:AT-93,Corkum:PP-93,%
Lewenstein:HH-94,Agostini:PA-04}.

Several extensions to the SAE~view of HHG have been investigated previously.
A two-electron scheme was considered that uses sequential double
ionization by an optical laser with a subsequent nonsequential
double recombination;
in helium this leads to a second plateau with about 12~orders of magnitude
lower yield than the primary HHG~plateau~\cite{Koval:ND-07}.
Two-color HHG (optical plus \XUV{}~light) has been studied in a
one-electron model~\cite{Ishikawa:PE-03} and with many-electron effects
included by a frequency-dependent polarization~\cite{Fleischer:GH-08};
the \XUV{}~radiation assists thereby in the ionization process leading
to an overall increased yield~\cite{Ishikawa:PE-03} and the emergence
of a new plateau~\cite{Fleischer:GH-08}, the latter, however,
at a much lower yield.
The above schemes suffer from tiny conversion efficiency beyond the
conventional HHG~cutoff (maximum photon energy).

\begin{figure}
  \begin{center}
    \includegraphics[clip,width=\hsize]{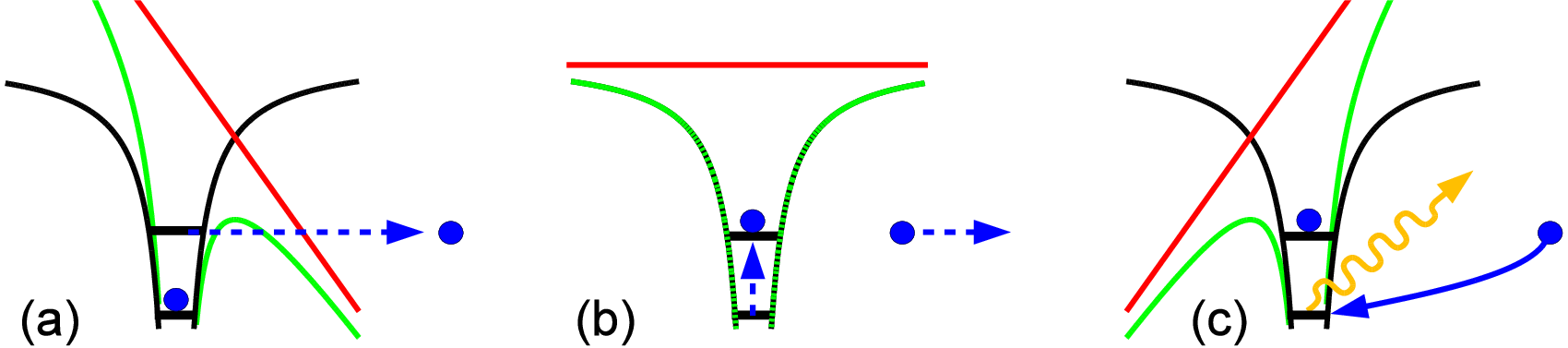}
    \caption{(Color online) Schematic of the three-step model for the
             HHG~process augmented by \XUV{}~excitation of a core electron.}
    \label{fig:3stepXray}
  \end{center}
\end{figure}

We propose an efficient two-electron scheme for a HHG~process manipulated by
intense \XUV{}~light from the newly constructed free electron
lasers~(FEL)---\eg, the Free Electron Laser in Hamburg~(FLASH).
Our principal idea is sketched in Fig.~\ref{fig:3stepXray}.
In the parlance of the three-step model~\cite{Schafer:AT-93,Corkum:PP-93},
HHG proceeds as follows:
(a)~the atomic valence is tunnel ionized;
(b)~the liberated electron propagates freely in the electric
field of the optical laser;
(c)~the direction of the optical laser field is reversed and the electron
is driven back to the ion and eventually recombines with it emitting
HHG~radiation.
The excursion time of the electron from the ion is approximately~$1 \U{fs}$
for typical $800 \U{nm}$~optical laser light.
During this time, one can manipulate the ion such that the returning
electron sees the altered ion as depicted in Fig.~\ref{fig:3stepXray}.
Then, the emitted HHG~radiation bears the signature of the change.
Perfectly suited for this modification during the propagation step is
\XUV{}~excitation of an inner-shell electron into the valence shell.
The recombination of the returning electron with the core hole leads
to a large increase of the energy of the emitted HHG~light as the energy
of the \XUV{}~photons~$\omega\I{X}$ is added
shifting the HHG~spectrum towards higher energies.
A prerequisite for this to work certainly is that the core hole is
not too short lived, \ie, it should not decay before the continuum
electron returns.

The spatial one-electron states of relevance to the problem are the valence
state~$\ket{a}$ and the core state~$\ket{c}$ of the closed-shell atom.
In strong-field approximation, continuum electrons are described by
free-electron states~$\ket{\vec k}$ for all~$\vec k \in
\mathbb R^3$~\cite{Lewenstein:HH-94}.
The associated level energies are~$E_a$, $E_c$, and $\frac{\vec k^2}{2}$.
We need to consider three different classes of two-electron basis
states to describe the two-electron dynamics:
first, the ground state of the two-electron system is given by the
Hartree product~$\ket{a} \otimes \ket{c}$;
second, the valence-ionized states with one electron in the continuum
and one electron in the core state are~$\ket{\vec k} \otimes \ket{c}$;
third, the core-ionized states with one electron in the continuum
and one electron in the valence state are~$\ket{\vec k} \otimes \ket{a}$.
We apply the three assumptions of Lewenstein~\etal~\cite{Lewenstein:HH-94}
in a somewhat modified way by considering also phenomenological decay
widths of the above three state: $\Gamma_0$ and $\Gamma_a$~to
account for losses due to ionization by the optical and \XUV{}~light
for~$\ket{a} \otimes \ket{c}$ and $\ket{\vec k} \otimes \ket{c}$,
respectively, and $\Gamma_c$ to represent losses from ionization
by the optical and \XUV{}~light and Auger decay of core holes
for~$\ket{\vec k} \otimes \ket{a}$ with radiative decay of the
core hole being safely neglected.
Further, the \XUV{}~light induces Rabi flopping in the two-level system
of~$\ket{\vec k} \otimes \ket{a}$ and $\ket{\vec k} \otimes \ket{c}$.

The two-electron Hamiltonian of the atom in two-color light
(optical laser and \XUV{}) reads~$\hat H = \hat H\I{A} + \hat H\I{L}
+ \hat H\I{X}$;
it consist of three parts: the atomic electronic structure~$\hat H\I{A}$,
the interaction with the optical laser~$\hat H\I{L}$, and the interaction
with the \XUV{}~light~$\hat H\I{X}$.
We construct~$\hat H$ mostly from tensorial products of the corresponding
one-particle Hamiltonians~$\hat h\I{A}$, $\hat h\I{L}$, and
$\hat h\I{X}$.
The interaction with the optical and \XUV{}~light is treated in
dipole approximation in length form~\cite{Buth:HO-up}.

We make the following ansatz for the two-electron wavepacket (in atomic units)
\begin{eqnarray}
  \label{eq:wavepacket}
  \textstyle \ket{\Psi, t} &=& \textstyle a(t) \, \euler^{-\frac{\imag}{2}
    \, (E_a + E_c
    - \omega\I{X}) \, t+\imag \, \Cal I\I{P} \, t} \ket{a} \otimes \ket{c} \\
  &&\textstyle {} + \Int_{\mathbb R^3} \bigl[ b_a(\vec k,t) \, \euler^{
    -\frac{\imag}{2} \, (E_a + E_c - \omega\I{X}) \, t + \imag \,
    \Cal I\I{P} \, t} \ket{\vec k} \otimes \ket{c} \nonumber \\
  &&\textstyle {} + b_c(\vec k,t) \, \euler^{-\frac{\imag}{2} \, (E_a + E_c +
    \omega\I{X}) \, t + \imag \, \Cal I\I{P} \, t} \ket{\vec k} \otimes
    \ket{a} \bigr] \differential^3 k \; , \nonumber
\end{eqnarray}
where we introduce a global phase factor based on~$\Cal I\I{P} =
-E_a + \frac{\delta}{2}$.
The detuning of the \XUV{}~photon energy from the energy difference
of the two ionic levels is~$\delta = E_a - E_c - \omega\I{X}$.
The index on the amplitudes~$b_a(\vec k,t)$ and $b_c(\vec k,t)$
indicates which orbital contains the hole.

We insert~$\ket{\Psi, t}$ into the time-dependent Schr\"odinger
equation and project onto the three classes of basis
states which yields equations of motion~(EOMs) for the involved coefficients.
We obtain the following EOM for the ground-state population
\begin{equation}
 \label{eq:groundepl}
 \dfrac{\differential}{\differential t} \, a(t) = -\frac{\Gamma_0}{2}
   \, a(t) - \imag \, \Int_{\mathbb R^3} b_a(\vec k,t) \, \bra{a}
   \hat h\I{L} \ket{\vec k} \differential^3 k \; .
\end{equation}
The other two EOMs are written as a vector equation, defining the
amplitudes~$\vec b(\vec k, t) = \bigl( b_a(\vec k, t),
b_c(\vec k, t) \bigr)\transpose$, the Rabi
frequency~$R\I{0X}$~\cite{Meystre:QO-99}
for continuous-wave~(CW) \XUV{}~light and the Rabi matrix
\begin{equation}
  \label{eq:RabiMat}
  \mat R = \left( \begin{matrix}
    -\delta - \imag \, \Gamma_a & R\I{0X} \\
    R\I{0X}                     & \delta - \imag \, \Gamma_c
  \end{matrix} \right) \; .
\end{equation}
This yields for a CW~optical laser electric field~$E\I{L}(t)$
oscillating with angular frequency~$\omega\I{L}$:
\begin{eqnarray}
 \label{eq:coupledAmps}
 \dfrac{\partial}{\partial t} \, \vec b(\vec k, t)
   &=& -\dfrac{\imag}{2} \, \bigl( \mat R + (\vec k^2
   + 2 \, \Cal I\I{P}) \, \unitmatrix \bigr) \, \vec b(\vec k, t)
   + E\I{L}(t) \\
 &&{} \times \dfrac{\partial}{\partial k_z}  \, \vec b(\vec k, t)
   - \imag \, a(t) \, E\I{L}(t) \, \bra{\vec k} \hat h\I{L} \ket{a}
   \, \binom{1}{0} \; . \nonumber
\end{eqnarray}
We change to the basis of \XUV{}-dressed states using the
eigenvectors~$\mat U$ and eigenvalues~$\lambda_+$, $\lambda_-$
of~$\mat R$.

To determine the HHG~spectrum, we solve Eq.~(\ref{eq:groundepl})
by neglecting the second term on the right-hand side as in
Ref.~\onlinecite{Lewenstein:HH-94}---its influence is included in~$\Gamma_0$,
$\Gamma_a$, and $\Gamma_c$~---and a constant \XUV{}~flux starting at~$t = 0$
and ending at~$t = T\I{P}$.
The HHG~spectrum is given by the Fourier
transform of~$\bra{\Psi_0,t} \hat D \ket{\Psi\I{c}, t}$,
where $\hat D$~is the two-electron electric dipole operator~\cite{Buth:HO-up},
$\ket{\Psi_0,t}$~is the ground-state part of the
wavepacket~(\ref{eq:wavepacket}) and $\ket{\Psi\I{c}, t}$~is
the continuum part, \ie,
\begin{eqnarray}
  \label{eq:dipcontinuous}
  \textstyle \tilde{\Cal D}(\Omega)
  &=& \textstyle -\imag \> \Sum_{\atopa{\scriptstyle i \in \{a, c\}}
    {\scriptstyle j \in \{+, -\}}} U_{ij} \, w_{j}
    \Int_0^{\infty} \sqrt{\tfrac{(-2 \pi
    \imag)^3}{\tau^3}} \> \euler^{-\imag \, F_{0,j}(\tau)} \nonumber \\
  &&\textstyle {} \times \Sum_{N = -\infty}^{\infty} \imag^N
    J_N\bigl(\frac{U\I{P}}{\omega\I{L}} \> C(\tau) \bigr)
    \; \euler^{\imag \, N \, \omega\I{L} \, \tau} \\
  &&\textstyle {} \times \Sum_{M = -\infty}^{\infty} \mathfrak b_{M-N,i}(\tau)
    \, h_{M,0,i}(\Omega, \tau) \differential \tau \; . \nonumber
\end{eqnarray}
Here, $\vec w = \mat U^{-1} \, \binom{1}{0}$, the ponderomotive
potential of the optical laser is $U\I{P}$, and $F_{0,j}(\tau)$, $C(\tau)$ are
defined as in Ref.~\onlinecite{Lewenstein:HH-94}
augmented by $\omega\I{L}$ and with $I\I{P}$
replaced by $\frac{\lambda_j}{2}+\Cal I\I{P}$. Further, $J_N$
are Bessel functions and $\mathfrak b_{M-N,i}(\tau)$~are the coefficients
defined in Eq.~(17) of Ref.~\onlinecite{Lewenstein:HH-94} for valence-
and core-hole recombination.
Further,
\begin{equation}
  \label{eq:lineshape}
  h_{M,N,i}(\Omega, \tau) = \dfrac{\euler^{\frac{\Gamma_0}{2} \, \tau} \,
    (1 - \euler^{-\Gamma_0 \, T\I{P} - \imag \, (\tilde\Omega_{M,N,i}
    - \Omega ) \, T\I{P}})}{\Gamma_0 + \imag \, ( \tilde\Omega_{M,N,i}
    - \Omega )} \; .
\end{equation}
Neglecting the factor~$\euler^{\frac{\Gamma_0}{2} \, \tau}$ for now,
we see that the~$h_{M,N,i}(\Omega, \tau)$ peak
at~$\Omega = \tilde\Omega_{M,N,i} = (2 \, (M+N) + \delta_{i,a}) \, \omega\I{L}
+ \delta_{i,c} \, \omega\I{X}$.
In other words, for~$i = a$ the peaks are at the positions of
the harmonics from optical laser-only HHG;
for~$i = c$, the harmonics are shifted by~$\omega\I{X}$ with respect to the
harmonics for~$i = a$ such that, in general, none of them coincides
with harmonics from~$i = a$.
The harmonic photon number spectrum~(HPNS) for a single atom---the probability
to find a photon with specified energy---along the $x$~axis is given by
\begin{equation}
  \dfrac{\differential^2 P(\Omega)}{\differential \Omega \differential
    \Omega\I{S}} = 4 \, \pi \, \Omega \, \varrho(\Omega) \>
    |\tilde{\Cal D}(\Omega)|^2 \; ,
\end{equation}
with the density of free-photon states~$\varrho(\Omega)$~\cite{Meystre:QO-99}
and the solid angle~$\Omega\I{S}$.

We apply our theory to krypton atoms.
%
%
The energy levels
%
%
are~$E_a = -14.0 \eV$ for~Kr$\,4p$~\cite{Yoon:RS-94} and
$E_c = -96.6 \eV$ for~Kr$\,3d$ with a radial dipole
transition matrix element of~$0.206 \U{Bohr}$~\cite{Buth:HO-up}.
The \XUV{}~light has the photon energy~$\omega\I{X} = E_a - E_c$.
The optical laser intensity is set to~$I\I{0L} = 3 \E{14} \U{\frac{W}{cm^2}}$
at a wavelength of~$800 \U{nm}$.
Both \XUV{} and optical light have a pulse duration
of~$T\I{P} = 3 \, \frac{2 \pi}{\omega\I{L}}$.
The experimental value for the decay width of Kr$\, 3d$~vacancies
%
%
is~$\Gamma\I{expt} = 88 \U{meV}$~\cite{Jurvansuu:IL-01}.
The decay widths due to \XUV{}~ionization of the atom and
the ion are obtained from the sum of the photoionization cross
sections of the energetically accessible
electrons~\cite{LANL:AP-00,Cowan:TA-81};
the ionization due to the optical laser is determined with
Ref.~\onlinecite{Krainov:IR-97}
and is larger than the width for \XUV~ionization and
Auger decay for the chosen parameters.
We find for an \XUV{}-intensity of~$I\I{0X} = 10^{13} \U{\frac{W}{cm^2}}$:
%
%
$\Gamma_0 = 280 \U{meV}$,
%
%
$\Gamma_a = 1.5 \U{meV}$, and
%
%
$\Gamma_c = 88 \U{meV}$,
and for an \XUV{}-intensity of~$I\I{0X} = 10^{16} \U{\frac{W}{cm^2}}$:
%
%
$\Gamma_0 = 450 \U{meV}$,
%
%
$\Gamma_a = 170 \U{meV}$, and
%
%
$\Gamma_c = 300 \U{meV}$.

\begin{figure}
  \begin{center}
    \includegraphics[clip,width=\hsize]{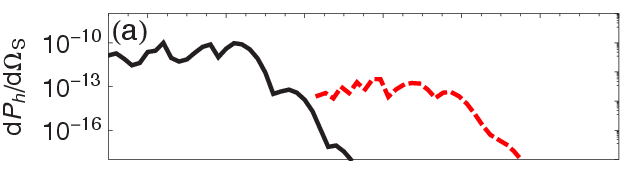}
    \includegraphics[clip,width=\hsize]{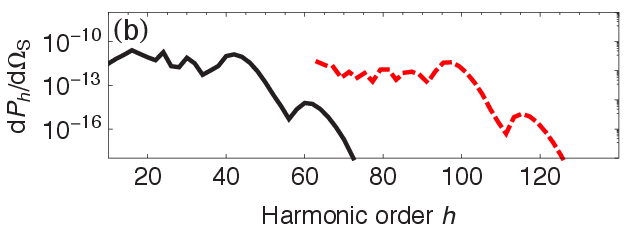}
    \caption{(Color online) Photon number of $h\,$th harmonic
             order for \XUV{}~intensities of (a)~$10^{13}
             \U{\frac{W}{cm^2}}$ and
             (b)~$10^{16} \U{\frac{W}{cm^2}}$.
             The black solid lines show the contribution from
             recombination with a valence hole whereas the red
             dashed lines correspond to recombination with a core hole.
             The lines represent harmonic strengths obtained by integrating
             over peaks in the HPNS.}
    \label{fig:HHGcompdepletion}
  \end{center}
\end{figure}

In Fig.~\ref{fig:HHGcompdepletion} we show the single-atom
HPNS of HHG which is modified by \XUV{}~light.
We find that the \XUV{}~excitation leads to two plateaus, one from
valence- and one from core-hole recombination which overlap slightly.
The width of the overlap can be tuned by changing the optical laser intensity
and interferences between both terms may occur.
Even in the case of a moderate \XUV{}~intensity
[Fig.~\ref{fig:HHGcompdepletion}a], the emission rate of HHG from core-hole
recombination is substantial.
This prediction can be interpreted in terms of an excitation (or even Rabi
flopping)~\cite{Meystre:QO-99} of the remaining electron after tunnel
ionization of an atom in the HHG~process [Fig.~\ref{fig:3stepXray}b].
This electron sits in a two-level system where the electron
is either in the valence or the core level.
The strength of the HHG~emission due to core-hole recombination is
roughly proportional to the population of the upper state around~$1 \U{fs}$.
For~$10^{13} \U{\frac{W}{cm^2}}$ the population
%
%
is~$\sim 0.001$ while it
%
%
is~$\sim 0.6$ for~$10^{16} \U{\frac{W}{cm^2}}$.
Finally, we need to realize that the dipole matrix element
for a recombination with a Kr$\,3d$~hole is not substantially different from
the one with a Kr$\,4p$~hole thus explaining the similar yield of both
contributions in Fig.~\ref{fig:HHGcompdepletion}b.

The \XUV{}~radiation from present-day FELs is generated with
the self-amplification of spontaneous emission~(SASE) principle;
it is fully transversally coherent but exhibits
only limited longitudinal coherence.
We find that the fluctuating phase does not destroy the
spectra~\cite{Buth:HO-up}.
Thus our interpretation of Fig.~\ref{fig:HHGcompdepletion} is not
invalidated when we relax the view assumed so far of entirely
coherent \XUV{}~light with constant amplitude.

In conclusion, we predict HHG~light from resonant excitation
of transient ions in HHG that allows insights into the physics
of core electrons and has various applications:
it allows one to generate isolated attosecond x-ray
pulses by ionizing atoms near the crests of a single-cycle optical
laser pulse and selecting the highest photon
energies by filtering~\cite{Goulielmakis:SC-08}.
This complements FEL-based strategies to generate
attosecond x~rays (Ref.~\onlinecite{Schneidmiller:LS-10} and References
therein).
Our scheme has the advantage that the attosecond pulses have a defined
phase-relation to the optical laser and that it can be employed at any FEL
with moderate cost and minimal impact on other experiments.
Further, our scheme has the potential to become
an \emph{in situ} probe of the dynamics of cations in strong optical fields
interacting with intense \XUV~light.
Namely, the HHG~spectra depend sensitively on the \XUV{}~pulse shape;
a reconstruction with frequency resolved optical
gating~(FROG)~\cite{Trebino:FR-02} may be possible---thus offering
the long-sought pulse characterization for SASE
\XUV{}~light---but requires further theoretical investigation.
Additionally, the emitted upshifted light due to core recombination
bears the signature of the \emph{core} orbital;
thus it can be used for ultrafast time-dependent chemical
imaging~\cite{Morishita:AR-08} involving inner shells which
is not feasible so far.
This allows one to extend such HHG-based methods to all orbitals
that couple to the transient valance vacancy by suitably
tuned \XUV{}~light.
Our findings are not restricted to krypton but HHG~spectra for resonant
excitation of $1s$~electrons in neon were successfully computed and
will be discussed in future work.

C.B.~and M.C.K.~were supported by a Marie Curie International Reintegration
Grant within the 7$^{\mathrm{th}}$~European Community Framework Program
(call identifier: FP7-PEOPLE-2010-RG, proposal No.~266551).
C.B.'s work was partially funded by the Office of Basic Energy Sciences,
Office of Science, U.S.~Department of Energy, under Contract
No.~DE-AC02-06CH11357.

\end{document}